# Thermodynamic investigation of an insulator irradiated by a low-energy electron beam

F. Pesty[a] and P. Garoche

Laboratoire de Physique des Solides, Université Paris-Sud, CNRS UMR 8502, Bât. 510, 91405 Orsay cedex, France

**Abstract.** The surface of an insulating material irradiated by a beam of low energy electrons charges positively if the yield of secondary electron is greater than unity. For such a dynamical equilibrium, the thermodynamic properties have been investigated by measuring the surface potential in response to a temperature oscillation of the material. It is shown that an oscillation amplitude of 0.4 K at 530 K induces an oscillation of the surface potential of about 0.5 volts. The frequency dependence indicates a monotonous decrease in the response with decreasing frequency, extrapolating to zero at zero frequency. We propose that this modification of the surface charge is driven by the temperature dependence of a gas of charged particles in equilibrium with the vacuum level.

**PACS.** 73.20.-r Electron states at surfaces and interfaces – 61.14.Hg Low-energy electron diffraction (LEED) and reflection high-energy electron diffraction (RHEED); 68.60.Dv Thermal stability; thermal effects

## 1 Introduction

The surface charging of insulating materials irradiated by an electronic bombardment is of considerable interest in many areas of physics and technology. It is related to important domains such as electron beam lithography and scanning electron microscopy [1]. For these fields the charging effect limits the resolution and is considered as a drawback to be suppressed or at least minimised. A considerable amount of effort has been devoted to the experimental and theoretical investigations of secondary electron emission from the surface of insulating materials. The problem has been treated by both classical electrostatic calculation [2] and Monte-Carlo simulation [3]. If we consider the evolution of the surface charging process as a function of the beam primary energy, three domains must be considered [4]. Below the first threshold value $E_{c1}$ the primary particle energy is too low, thus the number of secondary electrons per incident electron reemitted by the surface is smaller than one, so the surface charges negatively, more or less to the beam energy. In between $E_{c1}$ and $E_{c2}$, the second threshold energy, the yield of secondary electrons is greater than one and the surface charges positively [5,6]. Above $E_{c2}$, the surface charges negatively. Our present study will be limited to the second region, more precisely just above $E_{c1}$, where there is a good agreement between experiments and theoretical models [4] to claim that under electron irradiation the surface reaches an electrostatic equilibrium, with a positive charge located in the topmost surface atomic layer of the insulating material. Here we will focus on the thermodynamic properties of the positively-charged irradiated layer, in equilibrium with the beam. We will limit our investigation to the low-energy range, typically 60-100 eV, and unfocussed conditions, i.e., where the primary beam penetrates only the first atomic layers and the beam diameter is of the order of the millimetre. We are concerned by thermodynamic properties because we would like to know if such a dynamical electrostatic equilibrium is only the result of a mean static density of trapped charges, where individual contributions accumulate to form the surface potential, or if the formation of the positive layer gives rise, at least transitorily, to a two-dimensional gas of electrons in thermodynamic equilibrium. It is clear that the direct effect of a temperature change on individual particles with kinetic energy or potential energy of the order of 1 eV (typical for secondary electrons) can be safely ignored, however the situation is different for a free electron gas having a low density in thermal equilibrium. To address this question, we have investigated the electrostatic response of the irradiated insulator to a periodic thermal excitation of the order of a few tenths of Kelvin.

[a] E-mail : pesty@lps.u-psud.fr

## 2 Experiment

The thermodynamic measurement presented here is based on the simultaneous measurement of the sample temperature and of the sample surface charge. Let us first describe how the sample temperature is modulated. The thermal excitation is performed by infrared heating of the sample-holder. The latter acts as an electrostatic shield and is tied to the ground potential to avoid any direct electrostatic coupling with the insulating sample. A sine wave function of time for the thermal time evolution has been chosen in order to detect non-linearity and to use the thermal phase shift as a tool to analyse the thermal response. The heating device working in UHV has been described elsewhere [7]. It is able to raise the sample mean temperature between 400 and 600 K, and the heating power can be adjusted in order to produce a temperature oscillation whose amplitude, typically from 0.1 to 0.5 K, can be chosen independently of the frequency. The sample temperature is measured using a thermocouple, which is spot welded directly to the sample-holder.

Let us now describe how we measure the surface charge of the sample irradiated by a low-energy electron beam. Basically, we use a low-energy electron diffractometer that provides both the beam to irradiate the surface (typically 100 eV) and a phosphorus screen to image the diffracted beam. The diffraction pattern is monitored using a video camera. The video stream is digitised in phase with the thermal excitation. The time evolution of the diffraction pattern is numerically analysed in order to build a new image, whose pixel intensity $I$ is proportional to the oscillation amplitude for each pixel [8]. So, for a given pixel of row $i$ and column $j$, $I_{ij}(t)=\langle I_{ij}\rangle + \delta I_{ij}\sin(\omega t+\varphi_{ij})$, where the pulsation $\omega=2\pi f$ is that of the excitation. The frequency $f$ has been tuned between 0.06 and 0.23 Hz in the present study.

The method we use to evaluate the surface charge has been described elsewhere: it consists in analysing the oscillatory part of the diffraction pattern in response to a periodic thermal excitation [9]. Let us recall how it is used here to analyse the surface potential. If we consider the impact of a surface potential on the diffraction pattern, we must combine the diffraction process [10] with the beam deviation caused by the surface potential. The diffraction process quantises the component of the diffraction vector $K = k_{out} - k_{in}$ that is parallel to the surface, while the deviation conserves its perpendicular component. So the position of a given diffraction spot is absolutely not affected by a surface potential exhibiting an axial symmetry. However the diffraction occurs at a different energy so that the intensity of a given diffraction spot, which is a function of the electron wavelength, is indeed affected by the surface potential [9]. So any oscillation of the surface potential will lead to an oscillation of the spot intensities. In turn this spot intensity oscillation is used here to monitor accurately any variation of the surface potential under beam irradiation.

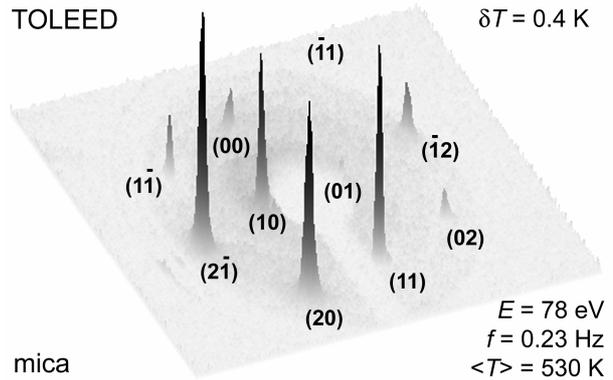

Fig. 1. Thermal response of the diffraction pattern. The 3D view displays the oscillatory part of the diffracted intensity, $\delta I$, in the case of the surface of muscovite mica. The grey level indicates the amplitude of the oscillation: darker shades of grey correspond to larger oscillations, while white shade indicates no oscillation. The maximum amplitude is about 3 grey levels, for the (2,-1) spot, representing about 8 % of the corresponding mean peak. Ten diffraction spots are indexed by their Miller indices. This oscillation pattern is the response of the insulating surface to a small temperature oscillation ($\delta T$ =0.4 K, peak).

The thermodynamic investigations have been performed on a muscovite mica sample (thickness 0.1 mm) cleaved in air and introduced into the analysis chamber within 10 min. The vacuum condition of the experiment corresponds to a pressure a few $10^{-10}$ Torr. The temperature oscillation $\delta T$ was between 0.1 and 0.5 K, around a mean temperature of 530 K. The working frequency was tuned between 0.06 and 0.23 Hz. The electron beam was adjusted to about 80 eV in order to be just above the threshold $E_{c1}$, in a region where the secondary electron yield is slightly greater than unity, so the surface charges positively. The beam current is kept constant and its value is fixed in the order of 0.5 µA (density about 50 µA/cm²).

## 3 Results

Figure 1 displays, in a 3D view, the image of the oscillation amplitude of the diffracted intensity in response to the thermal excitation. The diagram displays a shape very similar to a standard LEED (Low Energy Electron Diffraction) pattern, however it rather represents the oscillation of the intensity, $\delta I$. Surprisingly a small temperature oscillation, i.e., $\delta T$ =0.4 K, produces a significant evolution in the LEED intensity. For instance, the maximum amplitude $\delta I$ is found for the (2,-1) spot, and represents about 3 grey levels, i.e., 8 % of the corresponding $\langle I \rangle$. More generally, the intensity oscillation is about one order of magnitude below the intensity of the mean diffraction spot. It is worth noting that the background between the peaks is very low, because inelastic electrons generally display only a very weak energy dependence.

Note that each diffraction peak displays a specific response, because the oscillation amplitude is related to the derivative $\partial I/\partial E$ of a given Bragg spot intensity as a function of the beam primary energy [9]: $I = I(E + e<V> + e\delta V) \approx <I> + e\delta V(t)\partial I/\partial E$, where $V$ is the (positive) surface potential, the sum of a constant term and a sinusoidal one. So the oscillatory behaviour of the diffraction intensity is directly related to the oscillation of the surface potential.

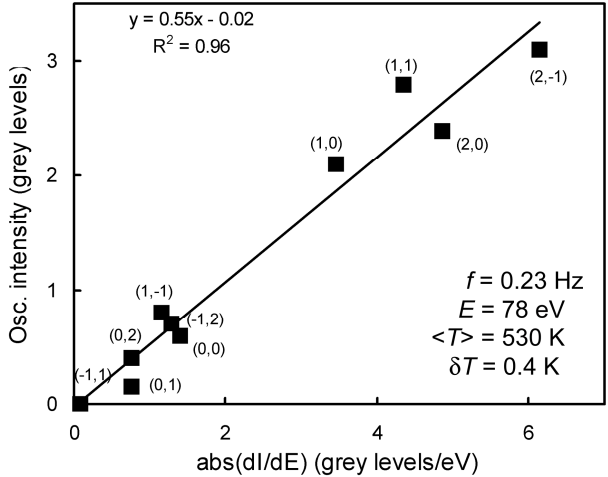

Fig. 2. Intensity oscillation as a function of the derivative of the diffraction intensity with respect to the energy. The oscillation amplitude of the diffraction spots of Fig. 1 is plotted versus the $\partial I/\partial E$ of each $(h,k)$ spot (in absolute value). The ten spots provide a common determination of the surface potential oscillation: from the slope of the fitted right line, we obtain a value of 0.55 V, produced by the 0.4 K temperature oscillation (peak values).

In order to accurately evaluate the amplitude of the latter, we can exploit all the Bragg spots of a single diffraction pattern. To this end, we have plotted in Figure 2 the intensity oscillation for the ten Bragg spots of Figure 1, as a function of the derivative of their intensity, in absolute value, with respect to the beam primary energy. We obtain a good linear relationship, from which we can calculate the oscillation of the surface potential: $\delta V = 0.55$ V. So a 0.4 K temperature oscillation causes a 0.55 V surface potential evolution, at the excitation frequency of 0.23 Hz. Moreover, the phase shift measured between the temperature oscillation and the intensity oscillation indicates that the surface potential increases as the temperature is increased.

In order to study the origin of this new thermo-electric effect, we have investigated how the amplitude of the surface potential oscillation evolves with the frequency of the thermal oscillation, for a fixed mean temperature, here $T=530$ K. The result is presented in Figure 3, where the oscillation amplitude of the surface potential is displayed as a function of the frequency, in the range 0.06 to 0.25 Hz for a constant temperature oscillation $\delta T$ of 0.4 K. The positive slope indicates that the thermo-electric effect increases with the thermal frequency and extrapolates to zero at zero frequency. So we are faced with a dynamical effect that shows up only when the system it driven out of thermal equilibrium. It must be noted that such a divergence of the response with the increasing frequency should reach a limit, however the latter is beyond the scope of this paper. In addition, the same thermo-electric effect has been observed in another insulating material: $SrTiO_3$ [11].

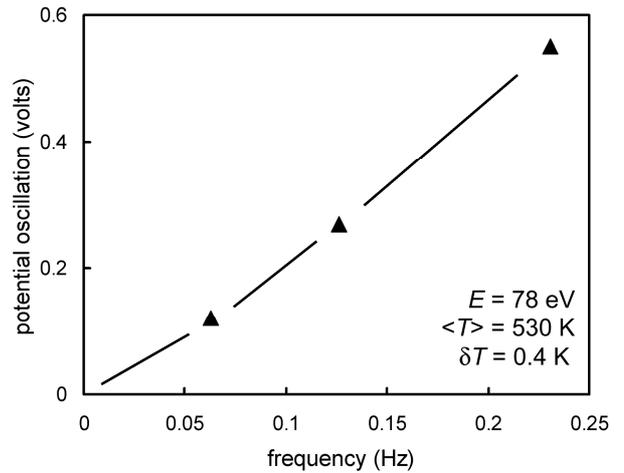

Fig. 3. Oscillation amplitude of the surface potential as a function of the excitation frequency. The amplitude of the temperature oscillation is kept around 0.4 K, for a mean surface temperature of 530 K. The response decreases as the frequency is lowered: no response is observed at zero frequency. Lines are guides for the eyes.

## 4 Discussion

Previous electrostatic investigations [2] have clearly shown that an irradiated surface accumulates the negative charges of the beam, and close to the surface, positive charges left by the emission of secondary electrons. In the unfocussed condition of our experiment, the irradiated volume is obviously two-dimensional, as depicted in Figure 4. This can be evaluated by comparing the penetration depth (about 1 nm at 80 eV) to the beam diameter (about 1 mm). In insulating materials all these charges are trapped. Their binding energy is well above the phonon energy (40 meV) so they can hardly respond to a small 0.4 K (0.03 meV) temperature evolution. Yet our thermodynamic investigation indicates a potential change in response to such a small thermal oscillation. Therefore, as proposed in the introduction, it appears that the large thermo-electric effect arises from an assembly of "free" charged particles in some form of equilibrium. So we are faced with a low-density two-dimensional gas of charged particles. We must consider a thermal equilibrium with the phonon temperature, and also an energy equilibrium with the vacuum level.

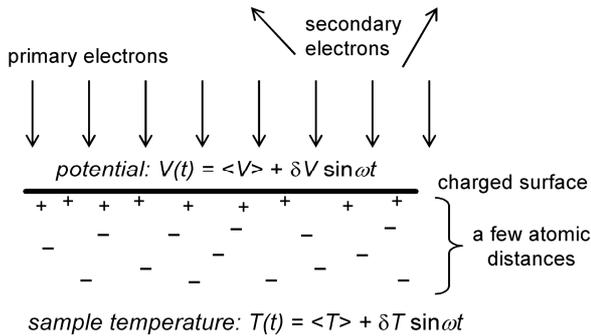

Fig. 4. Evolution of the electrostatic potential at the surface of an insulator. The sample is bombarded by a wide low-energy electron beam, establishing in the steady state a positively charged region. Located close to the topmost atomic layer, it results in a positive surface potential that tends to freeze the secondary electron emission, keeping dynamically the emission yield to unity. As the sample temperature is modulated, an oscillation of the surface potential is observed at the same frequency.

Another point is linked to the dynamics of the surface charge evolution: thanks to our phase shift information, we observe that the potential increases as the temperature is increased. This results from a change in charge density either as an increase in the number of positive charges or as a decrease in the number of negative charges. This last hypothesis is the most reasonable, if we are to consider both the better mobility of negative charges and the requirement of a thermal equilibrium, for which a temperature increase produces a reduction in the number of particles by evaporation towards the vacuum level. At very low frequency, the effective mobility of the positive charges is sufficient to cancel out the variation in the density of the electronic gas, so the thermo-electric effect tends to zero. These negative charges are probably located in the potential wells formed by the positive charges, but far enough from the surface (in the vacuum) to be free to move in two dimensions. It is well known that such "image states" have binding energies much smaller than typical work functions, and possibly of the same order of magnitude as the temperature [12]. It is clear that a precise description of this new phenomenon requires more experimental and theoretical investigations.

## 5 Conclusion

When a beam of low-energy electrons irradiates the surface of an insulating material, a dynamical electrostatic equilibrium is reached if the yield of secondary electron emission is greater than unity. The surface charges positively in order to adjust to unity the effective yield by trapping the secondary electrons. The thermodynamic properties of this dynamical electrostatic equilibrium have been investigated by oscillating the surface temperature of an insulator. A relationship between the surface charge and its temperature evolution has been discovered. A small oscillation of the surface temperature induces a significant modification of the surface charge, e.g., a modulation of 0.4 K around 530 K induces a 0.55V oscillation amplitude for the surface potential. This charge evolution can be roughly understood if we consider a low-density two-dimensional electron layer, which is maintained close to the surface by the positive layer. The frequency dependence shows that such an effect disappears at zero frequency. In fact at low frequency the slow positive carriers are able to cancel out the fast evolution of the electrons.

We thank C. Noguera, M. Bernheim and G. Blaise for fruitful discussions.